\documentclass[runningheads]{cl2emult}

\usepackage{makeidx}  
\usepackage{graphicx} 
\usepackage{subeqnar} 
\usepackage{multicol} 
\usepackage{cropmark} 
\usepackage{eso}      
\makeindex            

\begin{document}
\title*{Clustering of Very Red Galaxies in the Las Campanas IR Survey
}
\toctitle{Focusing of a Parallel Beam to Form a Point
\protect\newline in the Particle Deflection Plane}
%
%
\titlerunning{Las Campanas IR Survey}
%
\author{Patrick McCarthy\inst{1}
\and Ray Carlberg \inst{2}
\and Ron Marzke \inst{1,3}
\and Hsiao-Wen Chen \inst{1}
\and Andrew Firth \inst{4}
\and Richard McMahon \inst{4}
\and Jennifer Wilson \inst{1}
\and Eric Persson \inst{1}
\and Richard Ellis \inst{4,5}
\and Roberto Abraham \inst{2,4}
\and Ofer Lahav \inst{4}
\and Augustus Oemler \inst{1}
\and Chris Sabbey \inst{4}
\and Rachael Sommerville \inst{4}}
\authorrunning{McCarthy et al.}
%
%
\institute{ 1) Carnegie Observatories, 813 Santa Barbara St. Pasadena, CA,  USA\\
            2) Astronomy Dept., University of Toronto, Toronto, CANADA \\
            3) Astronomy \& Physics, San Francisco State Univ., San Francisco, CA, USA \\
            4) Institute of Astronomy, Cambridge University, Cambridge, UK  \\
            5) Astronomy Dept., Caltech, Pasadena, CA, USA \\
     }

\maketitle              

\begin{abstract}
\index{abstract} 
We report results from the first 1000 square arc-minutes of the Las Campanas IR
survey. We have imaged 1 square degree of high latitude sky in six distinct
fields to a 5$\sigma$ H-band depth of 20.5 (Vega). Optical imaging in the 
V,R,I,and z' bands allow us to select color subsets and photometric-redshift-defined
shells. We show that the angular clustering of faint red galaxies
($18 < H < 20.5, I - H > 3$) is an order of magnitude stronger than that of
the complete H-selected field sample. We employ three approaches to estimate $n(z)$ in order to invert
$w(\theta)$ to derive $r_0$. We find that our $n(z)$ is well described by a Gaussian with
$ <z> = 1.2, \sigma(z) = 0.15$. From this we derive a value for $r_0$ of $7^{+2}_{-1}$~ co-moving
$h^{-1}$Mpc at $<z> = 1.2$. This is a factor of $\sim 2$ larger than the clustering length
for Lyman break galaxies and is similar to the expectation for early type galaxies at this epoch.

\end{abstract}

\section{The Las Campanas IR Survey}

The LCIR survey was designed to measure the spatial clustering of early type galaxies 
at redshifts beyond 1 (Marzke et al. 1999; McCarthy et al. 1999). 
The motivations for this are numerous: clustering provides a means to
test competing variants of CDM and galaxy formation scenarios, it provides a path towards
matching low and high redshift galaxy populations and it provides an indirect measure of the
merging rate, a key part of the galaxy assembly process. The near-IR is the appropriate
window on massive evolved galaxies at large and intermediate redshifts. The red giant-dominated
spectral energy distributions of evolved galaxies when redshifted to $z \sim 1$ 
provide a distinctive spectral signature
against the rich foreground of faint blue galaxies. We used a variety of approaches to modeling
the colors of luminosities of the early type population in order to define the parameters
of our survey. Seeking to ensure a clean $5\sigma$ detection of clustering in the $1 < z < 2$
range we chose a filter set consisting of V,R,I,z',J and K$_s$ and a survey area of 1 square
degree. To comfortably reach K$_*$ at $z = 2$ requires a depth beyond K$_s = 20.5$, a quite ambitious
depth for a large area survey. In the present contribution we describe a ``pilot'' survey based
on H-band observations to a $5\sigma$ limit of 20.5 with complementary optical imaging over
one square degree. 

   The near-IR observations for this survey were made with the Cambridge Infrared Survey Instrument (CIRSI;
Beckett et al. 1999). This camera is built around a $2 \times 2$ array of Rockwell Hawaii I 
$1024 \times 1024$ HgCdTe detectors. The detectors are spaced by 90\% of a chip dimension in a checker-board
pattern. Filled mosaic images are built up by offsetting the telescope in four positions, covering the
gaps between the detectors. At the Cassegrain focus of the du Pont 2.5m telescope the size of a filled
mosaic, the basic unit of our survey, is $13^{'} \times 13^{'}$. 
To achieve our present depth of H$ = 20.5~ 5\sigma$ over the full square degree required
100,000 fully calibrated and processed $1024 \times 1024$ H-band frames.

   Our survey is spread over six fields at high galactic latitudes. This allows for year-round observing
and reduces sample variance effects in the clustering measurements. The fields are listed in Table one
and there is considerable overlap with a number of deep fields under discussion at this workshop. This
was intentional as our program is dependent on photometric redshifts and we wanted to be in fields where
there were abundant spectroscopic redshifts for use as calibrators. The tabulated areas are in
square arc-minutes and the depths are $5\sigma$ on the Vega scale.

\begin{table}
\centering
\caption{LCIR Survey Fields}
\renewcommand{\arraystretch}{1.4}
\setlength\tabcolsep{5pt}
\begin{tabular}{llllrl}
\hline\noalign{\smallskip}
Field &~~ $\alpha$ & ~$\delta$ & b & Area & H Depth \\
\noalign{\smallskip}
\hline
\noalign{\smallskip}
HDFS        &22 33 &-60& 49     &  1100   & 20.6         \\
NTT Deep    &12 05 &-07& 52     &   670   & 20.8         \\
CDFS        &03 32 &-27& 54     &   670   & 20.5         \\
NOAO S      &02 10 &-04& 60     &   670   & 20.5        \\
SA22        &22 00 &+00& 40     &   670   & 20.5        \\
IoA1511     &15 24 &+00& 44     &   670   & 20.5        \\

\hline
\end{tabular}
\label{Tab1a}
\end{table}

\section {Angular Clustering of Red Galaxies}
   
  We have assembled photometric catalogs for the first 1000 square arc-minutes of our survey, primarily
in the HDFS and NTT $12^h$ fields. From this catalog we have selected color subsets of the data,
focusing   on the reddest galaxies. In Figure 2 we show our color-magnitude and color-color
diagrams for 5200 H-selected galaxies.  The color range that we are focusing
our analysis on is defined by $I - H > 3$. These very red galaxies are not quite as extreme as the
canonical ``Extremely Red Objects'' (EROs), usually defined by R - K$> 6$, although our sample will
contain all of the EROs within our fields.
Selecting at wavelengths longer than $R$ gives us better sensitivity to early-type systems at $z \sim 1$ as we
are less likely to overlook old populations with low levels of on-going star formation. The abundance of
such systems is demonstrated by the wide range of $V - I$ colors displayed by our $I - H > 3$ sample
(Figure 1).
 
  We have computed the angular correlation function of our red sample. Our initial analysis in early
2000 was focused on a modest area ($13^{'} \times 39^{'}$) in the HDFS. We found a very strong 
clustering signal, 8 to 10 times stronger than that of the field. This result is very similar to
that published by Daddi et al. (2000) in their 800 square arc-minute survey to K = 18.8. Over the
past few months we have increased the size of our catalog of red galaxies
($I - H > 3, H < 20.5$)  and in Figure 3 we present the angular
clustering measurement for the first 1000 square arc-minutes of our sample. These data are drawn
from widely spaced fields and so provide a fair sample of structure within our color and
magnitude range. We recover the correct clustering scale for the complete field population of
$\sim 2^{''}$ and again find that the red population is 10 times more strongly clustered than
the population as a whole. Our best fit to $\theta_0$ for this sample is $12^{''}$. Similar to
the results reported by Daddi et al. (2000a,b,) we find that the clustering amplitude is a
strong function of the color. More extreme color-cuts produce larger, but poorly determined,
values for $\theta_0$, brighter magnitude cuts also produce larger clustering amplitudes, but again
the statistics are poor.

\begin{figure}
\centering
\includegraphics[width=1.0\textwidth]{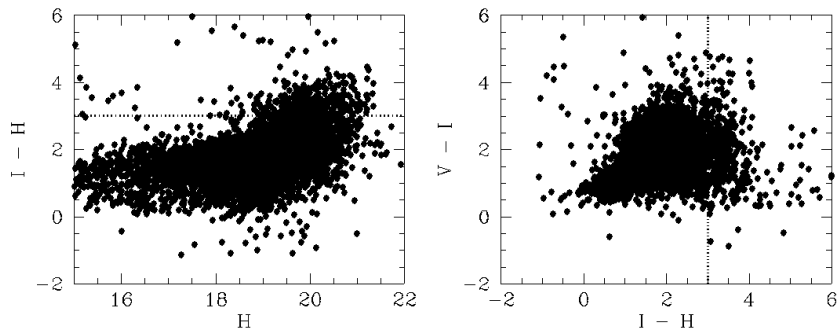}
\caption{The $I-H$ color magnitude and $V - I$ vs. $I - H$ color-color diagrams
for 5200 H selected objects in the first 1000 square arc-minutes of the LCIR survey. Our
$I - H > 3$ color selection threshold is shown with the dashed line.}
\label{fig1}
\end{figure}

\begin{figure}
\centering
\includegraphics[width=.8\textwidth]{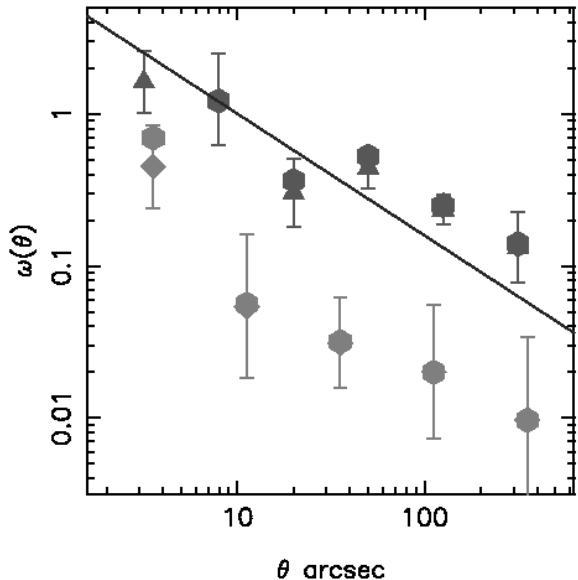}
\caption{The angular correlation function for red galaxies from the Las Campanas IR survey. The
lower points are for the full H$< 20.5$ sample, the upper points and curve are for the $I - H > 3$
subsample. The inversion has been done using both the Landy-Szalay and Hamilton algorithms and the
errors have been derived from a ``Jackknife'' analysis of the six sky areas that went into the
calculation. }
\label{fig1}
\end{figure}
\section {The Three Dimensional Clustering Length}

  The angular clustering scale can be converted into a three dimensional clustering length
if one knows the redshift distribution of the sample. We have estimated the $n(z)$ for our sample
using three approaches. The first involves modeling the color-magnitude and color-color
diagrams for our H-selected sample and the red population in particular. Using spectral evolution
models with conservative assumptions regarding the amount of passive evolution we infer that the
population defined by $I - H > 3, H < 20.5$ is largely confined to the $1 < z < 1.7$ range. A
more precise, but not necessarily more accurate, approach involves the use of photometric redshifts.
We find that the galaxies with $I - H > 3$ are
confined to the redshift range $1 < z < 1.5$. Lastly, we compare directly to the small area deep
redshift survey carried out by Cohen et al. (1999). Applying similar color and magnitude cuts to her
data again yields a redshift range from 1 to 1.5. We choose to model the $n(z)$ for our sample
as a Gaussian with $<z> = 1.2$ and $\sigma_z = 0.15$. The impact of moving $<z>$ by modest (e.g. $\sim 0.2$)
amounts on the derived value of $r_0$ is quite small; nearly all of the sensitivity is in the choice
of $\sigma_z$. This is intuitive as a thin redshift shell will produce a larger angular signal
for a given 3-D clustering length than will a broad redshift distribution. The thinness of our redshift
shell can be understood as arising from a combination of our color cut, which effectively removes the $z < 1$
foreground, and our magnitude limit, which places us well into the exponential portion of the luminosity
function by $z \sim 1.5$. These effects then give us fairly sharp upper and lower redshift cutoffs,
sharper than one would infer from simple pure luminosity evolution models of the color-magnitude
distribution.

   The value for $r_0$ that we derive is $7^{+2}_{-1}~h^{-1}$ co-moving Mpc. This is roughly twice
that of the Lyman break galaxies (Giavalisco, these proceedings) and is not much changed
from that of early type galaxies today. Our measured value for $r_0$ is also quite close to that
expected from hierarchical models for the formation of early type galaxies at early epochs (e.g. Kauffman
et al. 1999). The implications of this large clustering signal are not entirely clear and should be
approached with some caution. We know from sub-mm and spectroscopic studies (e.g. Graham and Dey 1997;
Cimatti et al. 1998; Dey et al. 1999) that some fraction of the very red galaxies are dusty star burst
systems. The strong clustering of very red galaxies suggests that this is a population for which
merging is important. Whether that merging has occurred recently and triggered massive starbursts or
is if it more indicative of the assembly of massive early type galaxies in dense environments
is not yet clear. There are several lines of evidence that suggest that the red population is dominated
by relatively dust free spheroids (e.g. Soifer et al. 1999; Yan et al. 1999). It seems plausible that
the ``EROs'', that is the most extreme objects in terms of colors and magnitudes
may contain a significant fraction of dusty systems, but that the bulk of the red galaxy population
with colors close to those expected from $z \sim 1$ ellipticals are indeed slowly evolving stellar 
systems. The evolution of the three dimensional clustering over the range from $z \sim 3$ to the
present offers one means of identifying and tracking such populations.

\clearpage
\addcontentsline{toc}{section}{Index}
\flushbottom
\printindex

\end{document}